# Tests of cosmic ray radiography for power industry applications

J. M. Durham[1*], E. Guardincerri, C. L. Morris[1], J. Bacon[1], J. Fabritius[1], S. Fellows[1] D. Poulson[1], K. Plaud-Ramos[1], and J. Renshaw[2]

[1]*Los Alamos National Laboratory, Los Alamos, NM 87544*

[2]*Electric Power Research Institute, Charlotte, NC 28262-8550*

24 March, 2015

**Abstract:** In this report, we assess muon multiple scattering tomography as a non-destructive inspection technique in several typical areas of interest to the nuclear power industry, including monitoring concrete degradation, gate valve conditions, and pipe wall thickness. This work is motivated by the need for radiographic methods that do not require the licensing, training, and safety controls of x-rays, and by the need to be able to penetrate considerable overburden to examine internal details of components that are otherwise inaccessible, with minimum impact on industrial operations. In some scenarios, we find that muon tomography may be an attractive alternative to more typical measurements.

LA-UR-15-22129

## Introduction

Naturally occurring background radiation from cosmic rays is present over the entire surface of the Earth. At sea level, most of the vertical charged particle cosmic ray flux consists of muons, which impinge upon the Earth's surface at a rate of $\sim 1/cm^2/min$ [1]. The relatively high energy of cosmic ray muons, their large mass (relative to electrons), and their lack of a strong interaction allow them to penetrate through dense objects, permitting investigation of the object's internal composition. The attenuation of cosmic rays as they pass through material has been used to study tunnel overburden [2] and pyramids [3], and has been more recently applied to measurements of volcanoes [4] and subway tunnel structure [5]. Multiple scattering of cosmic rays as they pass through objects has been shown to provide quantitative radiography for examining vehicles, transport containers, and cargo [6]- [7] , reinforced concrete columns [8], and the damaged cores of the Fukushima reactors [9].

At a typical cosmic ray muon energy of ~3-4 GeV, muons are minimum-ionizing particles with average energy loss per unit length $<-dE/dx>$ of only a few $MeV/(g/cm^2)$, making them a highly penetrating radiographic probe. Therefore objects can be inspected through considerable overburden from concrete, insulation, or other coverings, which can reduce operational downtime associated with removing insulation for inspection. In addition, since cosmic rays are natural background radiation, no artificial radiation sources are necessary. This technique avoids complications with regulatory requirements for ionizing radiation sources and has no potential to introduce additional dose to personnel.

---

[*] Corresponding author. Electronic address: *durham@lanl.gov*

To assess multiple scattering muon tomography as a non-destructive infrastructure inspection technique, muon multiple scattering imaging was performed on several representative test objects. All data described herein were obtained using the Los Alamos National Laboratory mini muon tracker (MMT). The MMT consists of two charged particle tracking detectors, one above and one below the test volume, that are each composed of 12 layers of 1.2 m long, 5 cm diameter drift tubes (see Fig. 1). The MMT was configured to measure vertical going cosmic rays with a test volume of approximately 100×100×60 $cm^3$. Objects under investigation are placed in between the two tracking modules, and the incoming and outgoing cosmic ray tracks are measured. The volume between the detectors is divided into voxels, and images are reconstructed based on the deflection of tracks passing through each voxel. For more details on the image reconstruction algorithms, see [10].

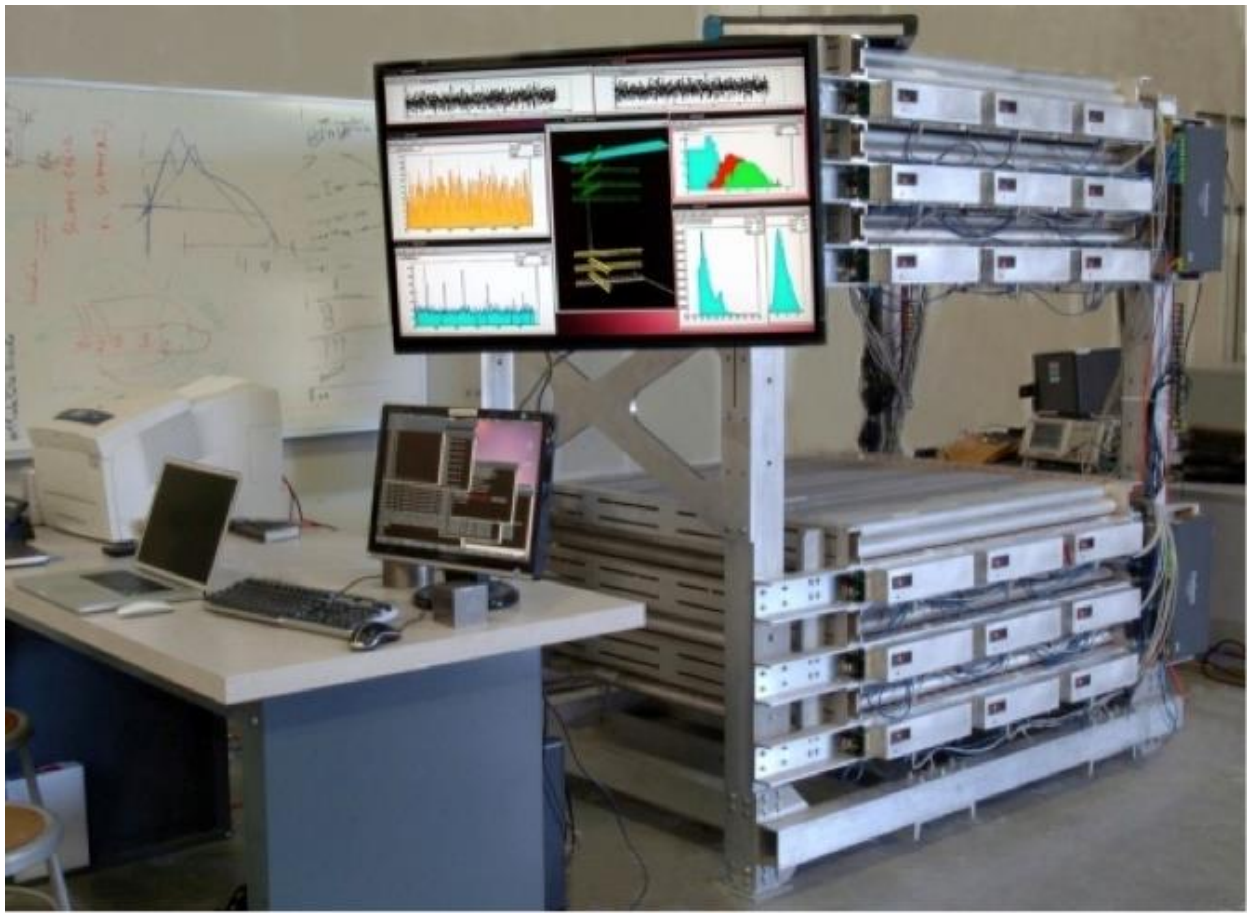

Figure 1: The Los Alamos Mini Muon Tracker. Two identical muon tracking modules are used to determine the scattering angle of muons that pass through objects placed in between them.

# Concrete test object

As concrete is a common construction material that is subject to erosion, techniques that can distinguish differences in concrete thickness without disturbing the object under investigation have a variety of applications. As a test of muon tomography, the first object to be evaluated was a set of four contiguous concrete tiles of thickness 7.6, 10.2, 12.7, and 20.3 cm, enclosed inside a plywood case (see Fig. 2).

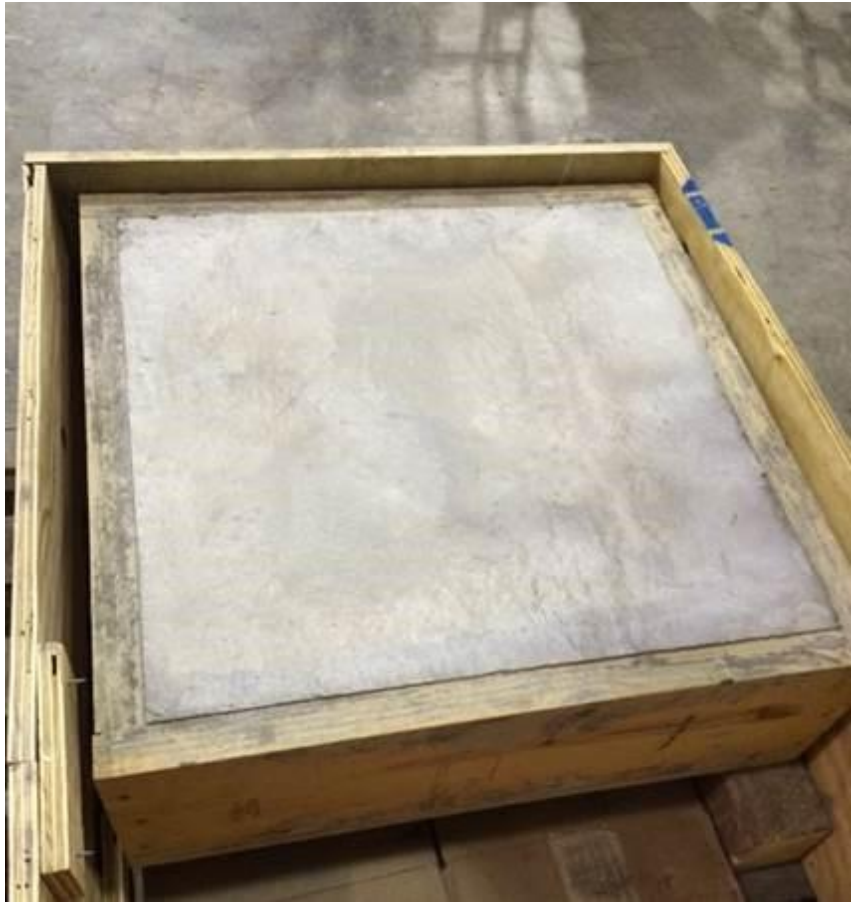

Figure 2: The concrete test object.

Data from cosmic ray tracks passing through the tiles was recorded for a total of 65 hours. Fig. 3 shows a series of stereo images of the concrete, moving through the concrete in 2 cm steps in the direction between the two detectors. The contrast of the images denotes the magnitude of the muon scattering in that voxel, which is related to the total radiation length experienced by the muon track. The object is observed to be constructed of 4 squares, each 30.5 cm on a side, and each with a different thickness.

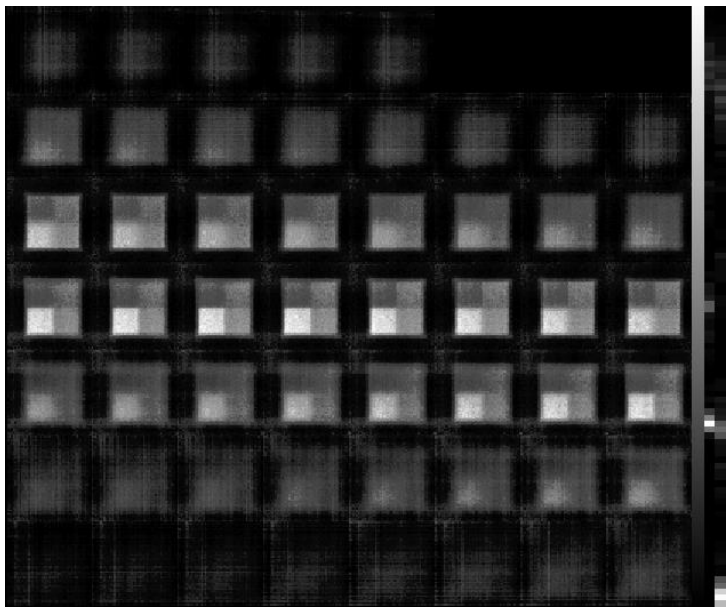

Figure 3) Stereo x-y slices of the concrete test object. Each panel is a step of 2 cm in the direction between the two muon tracking modules (z).

The data were analyzed using the multi group method described in [10]. This provides a quantitative measure of the object thickness in radiation length weighted areal density, $\rho_A/X_0$. These results are shown in Fig. 4. Clear differences are seen between each tile.

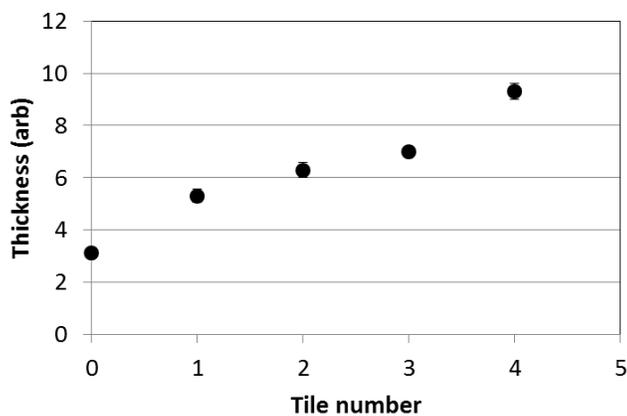

Figure 4) Concrete thickness for each of the tiles. Tile number 0 is an analysis of an empty spot in the detector volume, and represents the instrument background.

Fig. 5 shows the development of the image on an x-y slice through the center of the concrete as a function of exposure time. One hour exposure is sufficient to quantitatively distinguish the different concrete thicknesses.

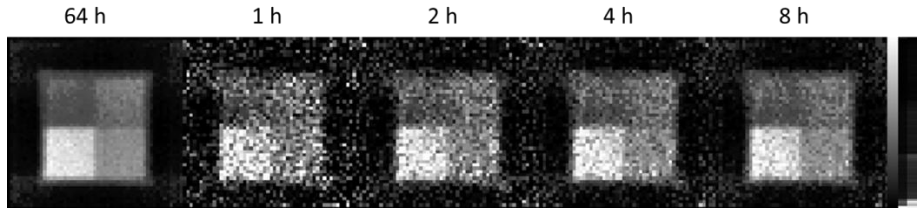

Figure 5) Image as a function of exposure time.

## Check Valve

A 7.6 cm check valve was imaged, with the goal of determining whether the position of the gate could be resolved and, if so, how long an exposure time was required. Tomographic images from a 4 hour exposure are shown in Fig. 6, for both closed and open gates. A side by side view of the slice at 94 cm is shown in Fig. 7, where the gate location is labeled.

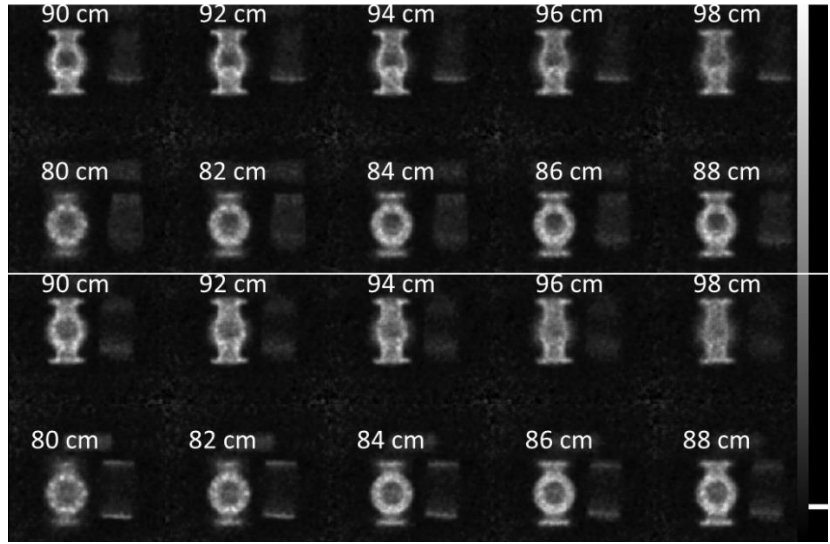

Figure 6) Tomographic reconstruction of the closed (bottom) and open (top) valve. The reconstructed images are shown as x-y slices as a function of depth between the two muon tracking modules (z). The object to the right of the valve is a section of pipe.

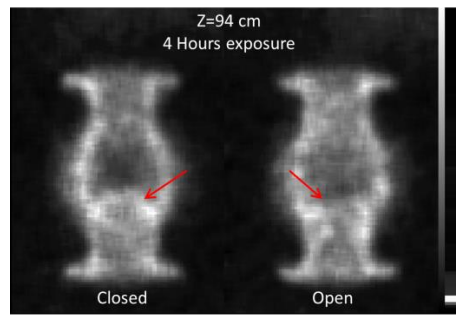

Figure 7) Expanded tomographic x-y slice at z=94 cm for closed (left) and open (right). The red arrow points to the region of the gate. A clear difference can be seen with a 4 hour exposure.

We analyzed 7 statistically independent one-hour-long runs by calculating the density in two 5 cm wide, 2 cm high patches of the image, one near the axis of rotation of the gate and one in the center of the opened gate. The mean and the standard deviation of the value of the radiation length weighted densities ($\rho_A/X_0$) were calculated for each of these regions. For both locations, the difference between open and closed is about a 3.5 standard deviation effect. The combination of the two gives a determination of the state of the gate at the 5 standard deviation level (the probability of an incorrect determination is $2.5\times10^{-7}$), proving that the gate condition can be accurately measured with a 4 hour exposure.

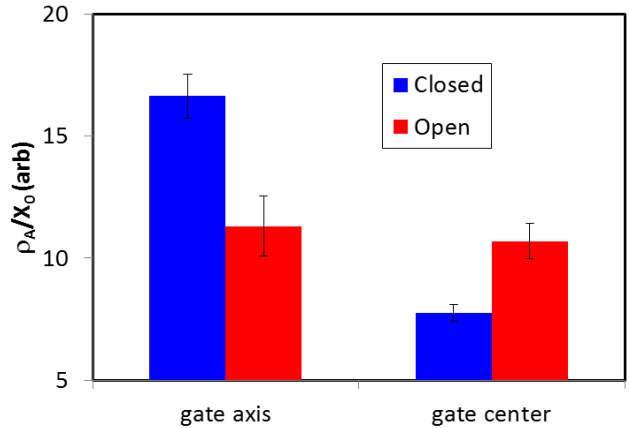

Figure 8) The mean (bars) and the standard deviation calculated for the densities measured on the gate axis (left) and the gate center (right) for the open (red) and closed positions (blue).

## Pipe wall thickness

Flow accelerated corrosion can lead to catastrophic failures of pipes and serious injuries to personnel [11]. Typical methods of inspection with ultrasound probes require insulation on pipes to be removed, rendering those pipe sections out of service for the duration of the measurement. Since muons are highly penetrating, pipes can be imaged through insulation layers while in use, minimizing impact on industrial operations.

To assess the ability of muon tomography to determine the thickness of pipes, we made measurements on lengths of stainless steel pipes with an outer diameter of 10 cm. Fig. 9 shows the time development of scattering data from two stainless steel pipes with different wall thicknesses, 4.1 mm and 13.3 mm. The outlines of the two pipes are distinguishable with only 15 minutes of exposure, and the contrast reflects the difference in areal density due to differing wall thickness. The data displayed are a center slice from the muon tomograph made with 1.5 cm wide overlapping pixels on 0.5 cm centers. At longer exposures the pipe walls can be observed in spite of the 1.5 cm pixel width.

We have analyzed 16 independent 1 hour long runs in order to determine the statistical precision for measuring the linear pipe density $\rho_L = \rho_V \pi \left(r_{otuside}^2 - r_{inside}^2\right)$, where $\rho_V$ is the volume density of the stainless steel, 7.8 g/cm³, $r_{ouside}$ is the outside radius of the pipe and $r_{inside}$ is the inside radius of the pipe. The wall thickness is $w=r_{outside}-r_{inside}$.

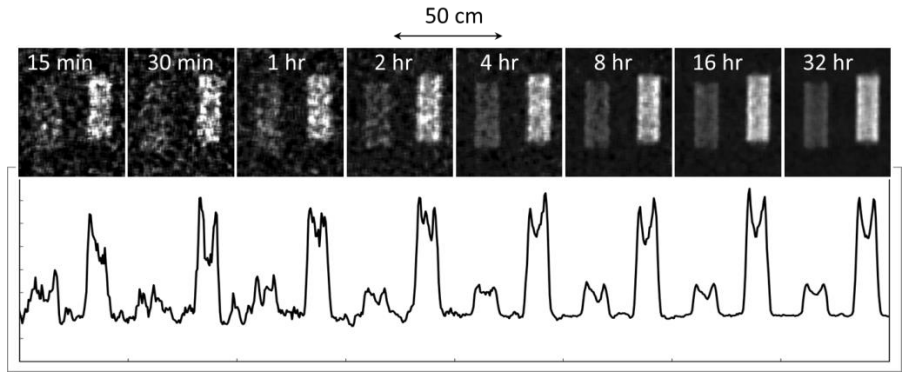

Figure 9) Images of scattering density showing the two stainless steel pipe sections. The line plot at the bottom is from a 10 cm wide slice.

Histograms of the results are shown in Figure 10. The standard deviation, SD, of the determination is 18% for the thin wall and 12% for the thick wall pipe. This variance arises in part from the background scattering signal induced by the ~4 cm of aluminum detector materials the muons must pass through. We estimate that by using thinner detectors and instrumenting four sides around an object can reduce the exposure times needed for a given measurement by more than a factor of 4.

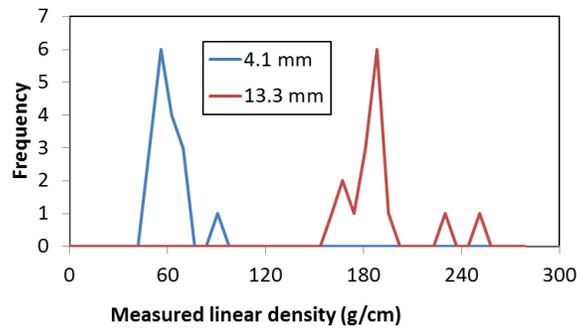

Figure 10) Histograms of the measured linear density of the two different thicknesses for 16 independent 1 hour long runs.

In Fig. 11 the measured vs the actual linear density of the pipes is shown. Here the averages of the 16 values are shown as the data points and the uncertainty, $\Delta = SD/\sqrt{16}$, is shown as the error bar. The data are linear within the several percent uncertainties. The slope of the line is 0.75, indicating that the absolute densities determined from the experiment are about 25% low. We attribute this discrepancy to an incomplete knowledge of the cosmic ray muon momentum spectrum, which is not directly measured with this technique. However, we note that the difference between the two pipes is readily apparent, making the technique useful for finding corroded sections.

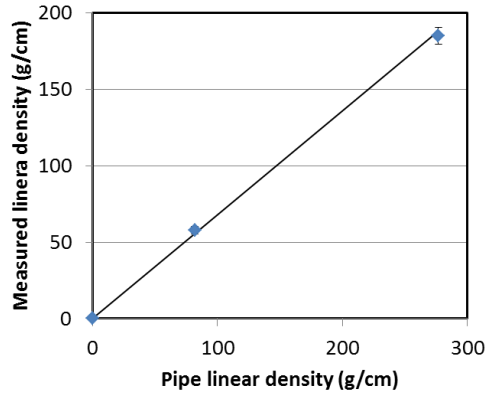

Figure 11) Measured vs actual pipe density. The line shows a linear fit to the data.

## Summary

We have presented data from tests aimed at demonstrating the capabilities of cosmic ray radiography for some typical inspection scenarios that may be encountered in the power industry. We have shown that cosmic rays can be used to gauge differences in concrete thickness, to image valves and determine their state, and to measure differences in pipe wall thicknesses. Cosmic rays provide a ubiquitous and safe method for imaging through significant overburden in situations where time intervals on the order of several hours are available for the imaging.  The penetrating ability of cosmic ray muons allows components embedded in concrete or insulation to be inspected while they are in use, avoiding disruption of industrial operations and the safety considerations that come with the use of artificial radiation sources.

While muon tomography does not provide the level of resolution that is found with ultrasound or x-ray measurements, it may still be useful in some contexts.  For example, it may be feasible to use muon radiography over lengths of pipe that are in use, and only resort to more invasive and precise methods of corrosion inspection when possible problem areas are found.

Currently, work is underway to develop new cosmic ray tracking detectors using tubes made of plastic or carbon fiber, rather than aluminum.  These new detectors will be significantly lighter, resulting in increased portability and ease of deployment.  In addition, the reduced materials budget of these new tubes will result in less muon scattering in the detector materials, which will improve the tracking precision and ultimately the resolution of reconstructed images.

## Acknowledgements

This work was performed under the auspices of the U.S. Department of Energy under Contract DE-AC5206NA25396. This work was support in part by Decision Sciences International Corporation and in part by Toshiba Corporation.

## References

[1] K. A. Oliver *et al.* (Particle Data Group), *Chin. Phys. C* **38**, 090001 (2014).

[2] E. P. George, *Commonwealth Engineer,* p. 455 (1955).

[3] L. E. Alvarez *et al.*, *Science,* **167** 3919 (1970).

[4] K. Nagamine *et al., Nucl. Instrum. and Meth. A* **365,** p. 585 (1995).

[5] S. Minato, *NDT and E International* **20**(4), p. 231 (1987).

[6] K. N. Borozdin *et al., Nature* **422**(6929): p. 277 (2003).

[7] C. L. Morris *et al., Science and Global Security* **16**(1-2), p. 37 (2008).

[8] H. K. M. Tanaka, *NDT and E International* **41**(3) p. 190 (2007).

[9] K. N. Borozdin *et al., Phys. Rev. Lett.* **109**(15), (2012).

[10] C. L. Morris *et al., AIP Advances* **2,** 042128 (2012).

[11] Electric Power Research Institute, Technical Report 1003619 (2003).